\documentclass[useAMS,usenatbib]{mn2e}
\usepackage{graphicx}

\newcommand{\CC}{$^{12}$CO}
\newcommand{\CCO}{$^{13}$CO}
\newcommand{\CDC}{$^{12}$C/\,$^{13}$C}

\newcommand{\Vt}{$V_{\rm t}$}

\newcommand{\Tef}{\mbox{$T_{\rm eff}$}}
\newcommand{\Msun}{\mbox{\,M$_\odot$}}

\newcommand{\vunit}{\mbox{\,km\,s$^{-1}$}}
\newcommand{\mic}{\mbox{$\,\mu$m}}

\newcommand{\ltsimeq}{\raisebox{-0.6ex}{$\,\stackrel
        {\raisebox{-.2ex}{$\textstyle <$}}{\sim}\,$}}
\newcommand{\gtsimeq}{\raisebox{-0.6ex}{$\,\stackrel
        {\raisebox{-.2ex}{$\textstyle >$}}{\sim}\,$}}

\newcommand{\RS}{RS~Oph}

\begin{document}

\title{First Overtone CO Bands in the Giant Component of RS~Ophiuchi: the
$^{12}$C/$^{13}$C Ratio in 2008}

\author[Ya. V. Pavlenko et al.]{Ya. V. Pavlenko$^{1}$\thanks
{E-mail:yp@mao.kiev.ua}, 
Charles E. Woodward$^2$,
M.T.Rushton$^{3}$, 
B. Kaminsky$^1$, 
A.Evans$^4$ \\
$^{1}$Main Astronomical Observatory, Academy of 
Sciences of the Ukraine, Golosiiv Woods, Kyiv-127, 03680 Ukraine \\
$^{2}$Department of Astronomy, School of Physics and 
Astronomy, 116 Church Street, S.~E., University of Minnesota, 
Minneapolis, MN 55455 \\ 
$^{3}$Jeremiah Horrocks Institute for Astrophysics and Supercomputing, University of Central 
Lancashire, Preston, Lancashire PR1 2HE   \\
$^{4}$Astrophysics Group, Keele University, Keele, 
Staffordshire, ST5 5BG, UK}

\date{}
\pagerange{\pageref{firstpage}--\pageref{lastpage}} \pubyear{2002}

\maketitle

\label{firstpage}

\begin{abstract}
We present an analysis of a high resolution ($R\simeq40\,000$) infrared 
spectrum of the RS~Oph secondary around the first overtone CO bands, 
obtained in 2008 May
on the Gemini South 8~m. The $^{12}$CO and $^{13}$CO bands are 
well-resolved, and we compute synthetic spectra to determine the \CDC\ 
ratio. We find \CDC\ $\simeq 16 \pm 3$, consistent with the 
interpretation of the secondary as red giant
which has evolved beyond the first drege-up phase of evolution.
\end{abstract}



\section{INTRODUCTION}

RS Ophiuchi is a recurrent nova (RN), an interacting binary system in 
which multiple nova outbursts have been observed; it has undergone 
eight known or suspected outbursts since 1898 \citep{anupama08}, the 
most recent occurring on 2006 February 12.83; we take this 
date as $t=0$. 

\RS\ consists of a white dwarf (WD) primary close to the Chandrasekhar 
limit, accreting material from a red giant (RG) secondary with a 
substantial wind \citep{fekel00}. The RN eruption is explained in terms of 
a thermonuclear runaway \citep[e.g.,][]{starrfield08,anupama08} following 
the accretion of sufficient mass by the WD to trigger ignition.
 If the mass accreted by the WD is greater than the mass ejected then the
mass of the WD must increase, so the ratio of mass accreted to mass 
ejected will determine whether or not \RS\ is a potential Type~Ia supernova 
progenitor \citep{sokoloski06,panagia06}. \citet{drake09}, modeling the 
X-ray absorption spectra observed with Chandra after the 2006 February 
outburst, argue that the mass ejected from the WD in the \RS\ system is 
$\sim 5 \times 10^{-6}~\rm{M}_{\odot}$. However, \citet{worters07} argue 
that the RG in the \RS\ system loses 
$\sim 10^{-8}~\rm{M}_{\odot}~\rm{yr}^{-1}$, while WD accretes at rates of 
$10^{-10} \ltsimeq $ \.{M}$_{\rm acc} \ltsimeq
10^{-9}~\rm{M}_{\odot}~\rm{yr}^{-1}$, 
suggesting that $\approx 2 \times 10^{-8}~\rm{M}_{\odot}$ of material is
accreted during the inter-outburst interval. 
Intercomparison of mass-loss rates from the RG, to the amount of material
ejected by the WD, can determine whether symbotic systems are candidate SN
progenitors \citep{panagia06}; however mass-loss rates are of course
difficult to pin down observationally.

\cite{pavlenko08_rsoph} showed that the secondary in \RS\ is a RG with 
\Tef = $4100\pm100$~K and solar metallicity, but with a deficit of carbon 
and an overabundance of nitrogen with respect to solar. \cite{rushton09} 
have modeled the spectral energy distribution of \RS\ in the near-IR
(1--5\mic) and found that \Tef\ is essentially constant over the period 2006
August -- 2008 July. Nitrogen overabundance, and carbon underabundance, 
is typical of ``normal'' red giants due to the conversion of C to N
in the CN-cycle of hydrogen burning and the subsequent dredge-up of the products
on the RGB \citep[e.g.,][]{smithlambert85,smithlambert86,smiljanic09}. 
Furthermore, the material accreting onto the WD from the M giant secondary in
the \RS\ system likely is N-enhanced \citep[see Fig.~14 of][]{ness09}. 
Interestingly, the C and N abundances for \RS\ found by 
\cite{pavlenko08_rsoph} are similar to those in the RG-WD binary (star 
\#69) in the open cluster IC\,4756 \citep{smiljanic09}. 

The \CDC\ isotopic ratio is of key importance for understanding the 
evolutionary status of RGs. At the beginning of the RG branch (RGB), the 
deepening convective envelope brings nuclear-processed matter from the 
core up to the surface of the star. As a result of the first dredge-up, 
the carbon isotopic ratio decreases up to a factor $\sim20$, depending on 
stellar mass and metallicty \citep{charbonnel94}, and standard models of 
stellar evolution predict no change in \CDC\ thereafter. 
However, observations of evolved RGs with progenitor mass 
$\ltsimeq2$\Msun\ indicate that the \CDC\ ratio continues to 
decrease after the completion of the first dredge-up 
event \citep[][ and references therein]{smith02,tautvaisiene,smiljanic09}. 
Most probably further mixing, or some non-standard mixing mechanisms, 
work in low mass stars as they are ascending the 
RGB \citep{charbonnel94,charbonnel95}. Anti-correlation of 
the carbon and nitrogen abundances, as is seen in 
\RS\ \citep{pavlenko08_rsoph}, is well known for a variety of RGs 
with a range of metallicities \citep[][ and references therein]{smith05}.

The \CDC\ ratio is also of fundamental importance in understanding the 
nature of the secondary of RNe, as it is a reflection of the relative 
importance of $^{12}$C, $^{13}$C processing and dredge-up.
 Following the 1985 eruption, \cite{evans85} estimated a \CDC\ ratio of 
$\sim10$ for the RG in \RS\ on the basis of a low resolution 
infrared (IR) spectrum, while \cite{scott94} found that the strength 
of the first overtone CO bands declined by a factor $\sim10$ over the 
first 7~years after the eruption \citep[see also][]{harrison93}.

In this paper, we discuss Phoenix echelle observations of the first overtone 
vibration-rotation features of CO, obtained in May 2008, some $831$ days  after
the 2006 outburst, to determine the \CDC\ ratio in the atmosphere of  the \RS\
secondary. 

\section{OBSERVATIONS AND REDUCTION}

High-resolution $H\!K\!M$-band spectral observations of \RS\ were 
carried out on the Gemini-South 8~m telescope using the 
Phoenix spectrograph \citep{hinkle03} during queue observing runs on 
the nights of 23 and 24 May 2008~UT (days 830.7, 831.7 from the 2006 
eruption), as part of our Gemini program GS-2008A-Q47. Defining zero 
orbital phase as maximum RG radial velocity, the observations discussed here
were obtained at binary phase 0.78, 
when approximately 88\% \citep[assuming the adopted inclination
$\sim40^\circ$;][]{fekel00,ribeiro09}
of the visible photosphere of the RG was illuminated by the WD.

Observations of \RS\ (as well as the telluric standard HR6378 
$= \eta~\mbox{Oph} \left[\rm{A2~V}, V = 2.43\right]$ at a comparable 
airmass) were obtained by nodding the target at multiple positions 
along the long axis of the $14\arcsec \times 0.34\arcsec$ slit 
\citep[e.g.,][]{smith02}. Observations of the telluric standard were 
conducted immediately before and after observations of \RS, with 
flats and darks being obtained at the conclusion of each observational 
sequence comprising a given grating and order-sorting filter 
combination. With our instrument configuration, the Phoenix 
spectrograph produces single-order echelle spectra with a resolution 
of $R = \lambda/\Delta\lambda \approx 40\,000$ (corresponding to a 
resolution element of $\sim 4$ pixels). Individual spectra were 
reduced and combined using standard IR data reduction techniques, as 
well as those unique to Phoenix \citep{hinkle03, 
hinkle07}. Table~\ref{table:_obstab} summarizes the observational data 
discussed herein. 

The echelle spectra discussed here covered the $2.32-2.33$~\mic\ and 
$2.36-2.38$~\mic\ spectral regions; we refer to these as spectra ``B'' 
and ``R'' respectively. Both spectra are affected by curvature of the 
echelle orders. We reduce the intensities of spectral features in most 
regions using the theoretical spectrum continuum as reference points. 
The effects are more pronounced in the ``B'' spectral region (see 
Fig.~\ref{fig:_ident}) and we work with the spectra  in terms of residual 
fluxes. We cannot determine the real continuum due to noise at the 
minimum fluxes. Therefore, we carried out fits to the observed 
spectrum using different levels of the continuum. These values of the 
virtual continuum we specify as pseudo-continuum. The results of the 
fits depends on the pseudo-continuum level, as discussed in section 
\ref{_r}. 

\section{ANALYSIS PROCEDURE}

\subsection{\textit{Spectral synthesis}}

We have used the technique of synthetic spectra to carry out our 
analysis of the IR spectrum of \RS; details may be found in 
\cite{pavlenko04,pavlenko08_rsoph}. We computed the spectra within the 
classical framework, assuming LTE, plane-parallel media, and no sinks 
or sources of energy in the atmosphere; transfer of energy is provided 
by the radiation field and by convection. Strictly speaking, these 
assumptions are not completely valid in the atmosphere of the RG in 
\RS; however, this approach allows us to describe the IR spectral 
energy distribution of \RS\ over a wide range of wavelengths, and we 
consider that none of our assumptions is of crucial importance in the 
spectrum formation processes. 

{\bf In the most general case, any determination of abundance depends on many
adopted parameters: \Tef, $\log{g}$ and \Vt, and the abundance of other
elements. In this paper we solve a restricted task using fits to 
the two observed narrow spectral regions in the inrared. Therefore,}  
to reduce the number of fitted parameters in the minimization
procedure, we adopt for the secondary atmosphere \Tef = 4200~K and $\log{g}
=0.0$, obtained from fitting near-IR low resolution spectra
\citep{pavlenko08_rsoph}.  We then fix the abundances of carbon and nitrogen at 
$\log{\mbox{$N$(C)}}=-4.2$ and $\log{\mbox{$N$(N)}}=-3.2$ on the scale 
$\Sigma N_{i} = 1$, {\bf as also determined from fitting the low resolution
spectra by \cite{pavlenko08_rsoph}}; other abundances are taken to be solar.

 We assume that the observed spectra are broadened by macroturbulence
in the RG atmosphere, and by instrumental broadening; both are implemented in
the modelling by smoothing the computed spectra with a gaussian of FWHM 
$\sim$1.9\AA, which is a reasonable approach for fitting stellar spectra. In
fact, the appropriate values of FWHM were determined by our minimisation
procedure fitting to observed spectrum \citep[see][]{pavlenko04}.

To determine the isotopic ratio \CDC\ we follow conventional
procedures. Namely, we define $x$ as the fraction of \CC\ molecules, i.e. $x =
N(\mbox{\CC})/(N(\mbox{\CC})+N(\mbox{\CCO})$, where $N(\mbox{M})$ is the
number density of the molecule M. Then, the fraction of \CCO\ molecules is 
clearly $N(\mbox{\CCO})/(N(\mbox{\CC})+N(\mbox{\CCO})) = 1- x$. In this
approach we operate with two parameters, i.e. the abundance of carbon, $\log
N(\mbox{C})$ {\bf (as specified above)} and the isotopic ratio $\mbox{\CDC} =
N(\mbox{\CC})/N(\mbox{\CCO})=x/(1-x)$. {\bf Then all synthetic spectra are
computed for the \CC\ and \CCO\ molecular densities, determined by taking into
account the input isotopic ratio \CDC.}

\begin{figure}
\includegraphics [width=80mm]{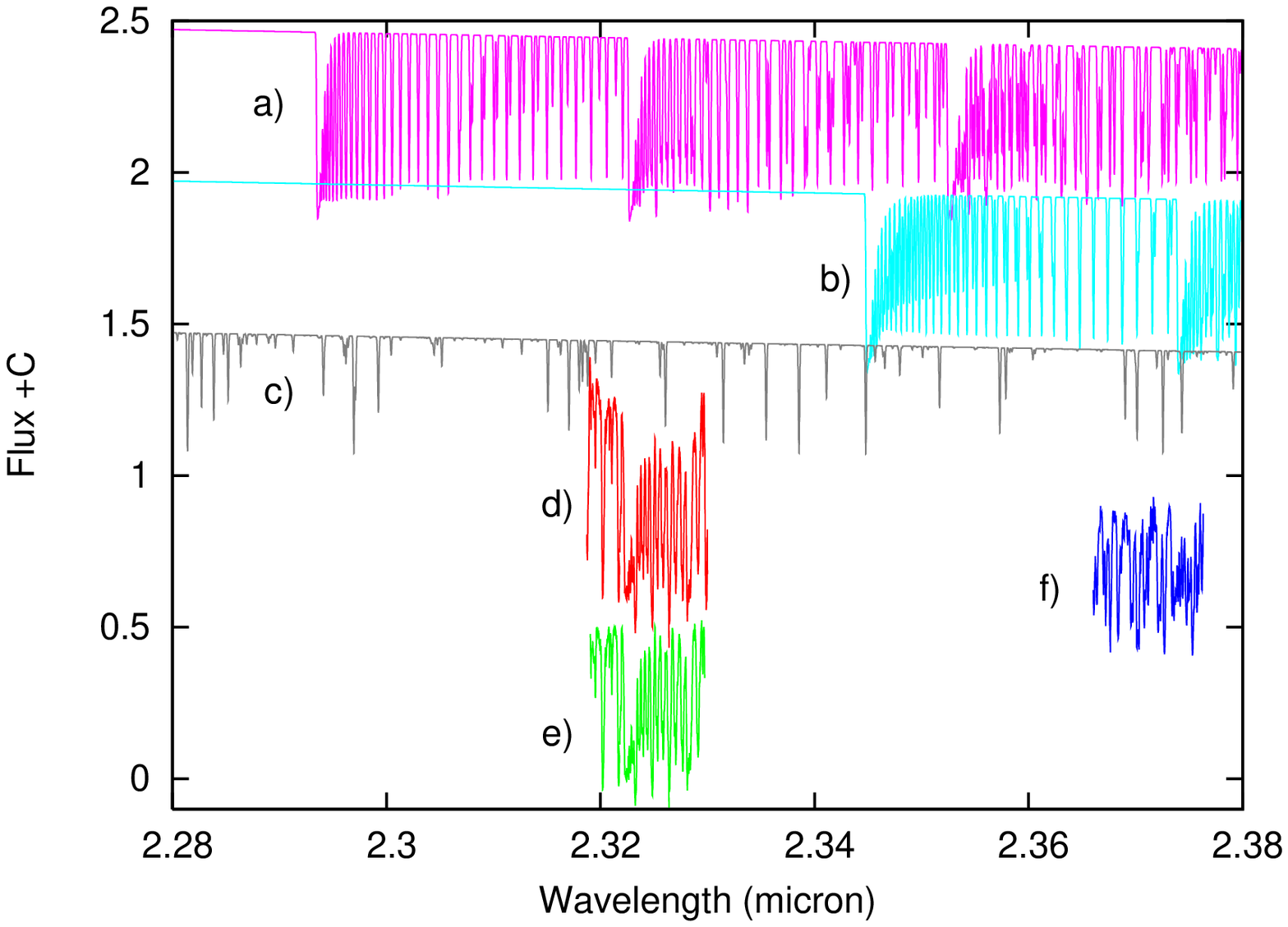}
\caption{Theoretical spectra of different species in 
our spectral region, convolved by a Gaussian with a half-width 
of $\times10^{-8}$\mic. Synthetic spectra were computed for a model 
atmosphere having $T_{\rm eff}=4000, \log{g}=0, \mbox{[Fe/H]}=0.0$ for 
(a)~\CC\, (b)~\CCO\, and (c) the VALD line lists. Observed spectra ($d, 
e, f$) in both spectral regions are also shown. The ``B'' spectrum is 
shown ``as is'' (d), and after echelle order curvature reduction (e). 
The flux scale is arbitrary, and all spectra are shifted vertically to 
simplify the plot.
\label{fig:_ident}
} 
\end{figure}

The ``B'' spectrum contains no \CCO\ lines, only \CC\ 
(Fig.~\ref{fig:_ident}). {\bf Again, we note that, in the most general case,
the intensities of these lines depend on both log N(C) and \Vt.} 
We used this spectrum to determine the 
microturbulent velocity $V_{\rm t}$ {\bf for the adopted 
value of log N(C)} ~in the atmosphere of the RG. The ``R'' 
spectrum contains both \CC\ and \CCO\ lines; we used this spectrum to 
determine the carbon isotopic ratio \CDC.  

\begin{table}
\caption{GEMINI PHOENIX OBSERVATIONAL SUMMARY\label{table:_obstab}}
\begin{tabular}{lccccc}
    
\hline
          &   &               &          &          \\ 
Date & Filter & $\lambda$ range & Exposure & Airmass \\
(2008 UT) &   & ($\mu$m)        &(s)       &         \\
\hline
          &   &               &          &          \\ 

23 May& K4308 & $2.32-2.33$ & 1650  & 1.18 \\
24 May& K4220 & $2.36-2.38$ & 1870 &  1.12  \\

\hline\hline
\end{tabular}
\end{table}

A set of continuum opacity sources from ATLAS9 \citep{kurucz93} 
with some additional amendments \citep{pavlenko04} was used in our 
modeling, as well as the VALD atomic line list \citep{kupka99}, and 
line lists of \CC\ and \CCO\ from \citet{goorvitch94}. The quality of 
the \citeauthor{goorvitch94} data is demonstrated by the fact that 
\cite{pavlenko08} recently fitted theoretical spectra to a high 
signal-to-noise 2.3\mic\ spectrum of Arcturus, giving excellent fits 
to the CO bands. Water vapor lines were not included because numerical 
experiments with the \cite{barber06} H$_2$O line list 
showed that their contribution to the total opacity in the
wavelength range under consideration is negligible. We adopt the Voigt
profile $H(a,v)$ for the shape of each line;  here  
$a = (\gamma_2 + \gamma_4 + \gamma_6)/(4\times\pi\times\Delta\nu_D),
v = \Delta\nu/\Delta\nu_D$ \citep[see][ for details]{gray}.
Damping constants $\gamma_2$, $\gamma_4$, and $\gamma_6$ 
are respectively due to
the natural, van der Waals and Stark broadening of atomic and molecular lines, 
computed using data from various databases \citep{kurucz93,kupka99}, or 
the approximation of \cite{unsold55}. Natural and van der Waals broadening
dominate in the case of RS~Oph, due to the low pressure and electron density in
the RG atmosphere. We note that 

\vspace{-2mm}

\begin{itemize}
\item[{--}] molecular lines are less sensitive to pressure broadening, 
because they form in low density regions of the RG atmosphere;

\item[{--}] thermal velocities of the CO molecule are lower than those of
C and O atoms due to its higher mass. For the CO molecular lines, broadening by
microturbulence dominates, especially in the low temperature regime; we
determine the microturbulence velocity in Section~\ref{_r}.
\item[{--}] the wings of molecular lines are blended in the spectra of the RG.
\end{itemize}

The contributions  of the various molecular and atomic species to the total
opacity in the  2.28--2.39\mic\ spectral region are shown in
Fig.~\ref{fig:_ident}, from  which we see that absorption by CO dominates. 

Full details of our procedure of fitting theoretical fluxes to observed 
SEDs may be found in \cite{pavlenko04}.

\subsection{\textit{The effect of circumstellar dust}}

Dust is present in the \RS\ system \citep{evans07, vanloon08, woodward08},
emitting primarily at $\lambda \gtsimeq 10$\mic. However \cite{rushton09} have shown
that there is also evidence for dust emission at shorter  ($\gtsimeq5$\mic)
wavelengths, and that emission by the dust envelope may be variable.

The properties of the dust shell around \RS\ have been discussed by 
\cite{evans07}, \cite{vanloon08} and \cite{woodward08}. 
\cite{pavlenko08_rsoph} and \cite{rushton09} have shown that the 
2\mic\ spectral region of \RS\ is not affected by emission from the 
hot ($\sim500-600$~K) dust known to currently envelop \RS, which
is likely a permanent feature of the system. As 
this dust emission becomes prominent longward of 3\mic, we are 
justified in carrying out our analysis for the dust-free case. 

\section{RESULTS}
\label{_r}

 In principle the \CDC\ isotopic ratio can be determined from any pair of the 
unblended lines of \CC\ and \CCO. However, in our spectra these lines are
blended and we are unable to distinguish the \CC\ lines from the weaker 
\CCO\ lines at our spectral resolution. 
Absorption of both species form common blends in the observed and modelled 
spectra. Only the band heads of the CO system can be clearly indentified 
in the spectrum, formed by saturated lines which have rather weak
dependence on the CO abundances. Moreover, they form in the outermost layers
of RG atmosphere which cannot be modelled in the framework of classical 
approach.
%
Non-saturated lines, especially those of intermediate strength, are a 
better choice for the abundance determination; however they are 
sensitive to the microturbulent velocity.

We determined the microturbulent velocity from the fits to the ``B'' 
spectrum ($2.32-2.33$\mic), which is formed completely by 
absorption of \CC\ lines. The spectrum shows a strong dependence on 
\Vt\ (Fig. \ref{fig:_vt}). The minimization procedure for the 
parameter $S$ (see below) gives a definite 
value for the microturbulent velocity of \Vt~=~$3.0\pm0.5$\vunit\ 
(close to the value assumed in \citealt{pavlenko08_rsoph}). In what 
follows, we compute spectra with \Vt~=~3\vunit\ to fit the ``R'' 
spectrum and determine the \CDC\ ratio in \RS. 

\begin{figure*}
\includegraphics [width=88mm]{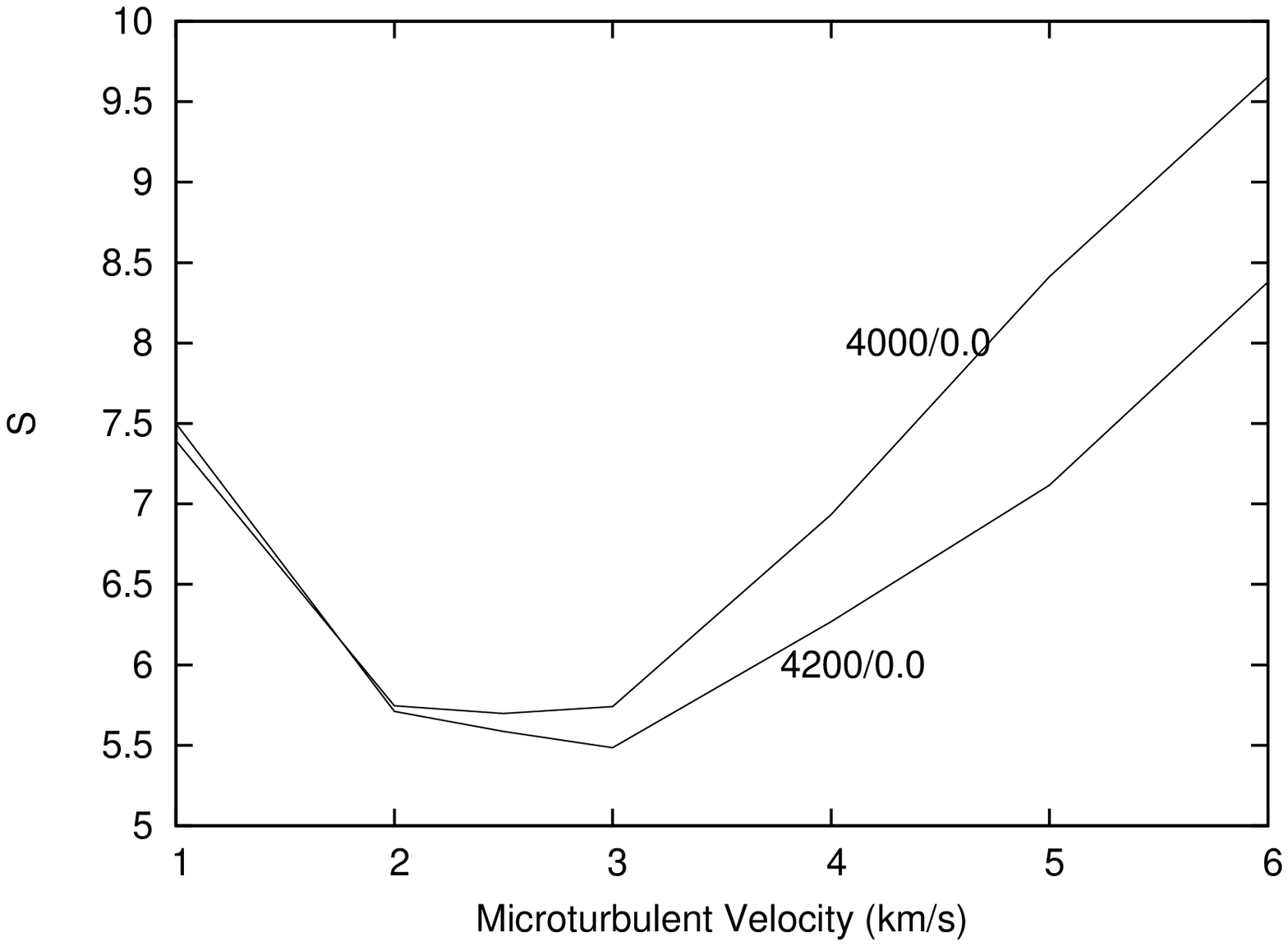}
\includegraphics [width=88mm]{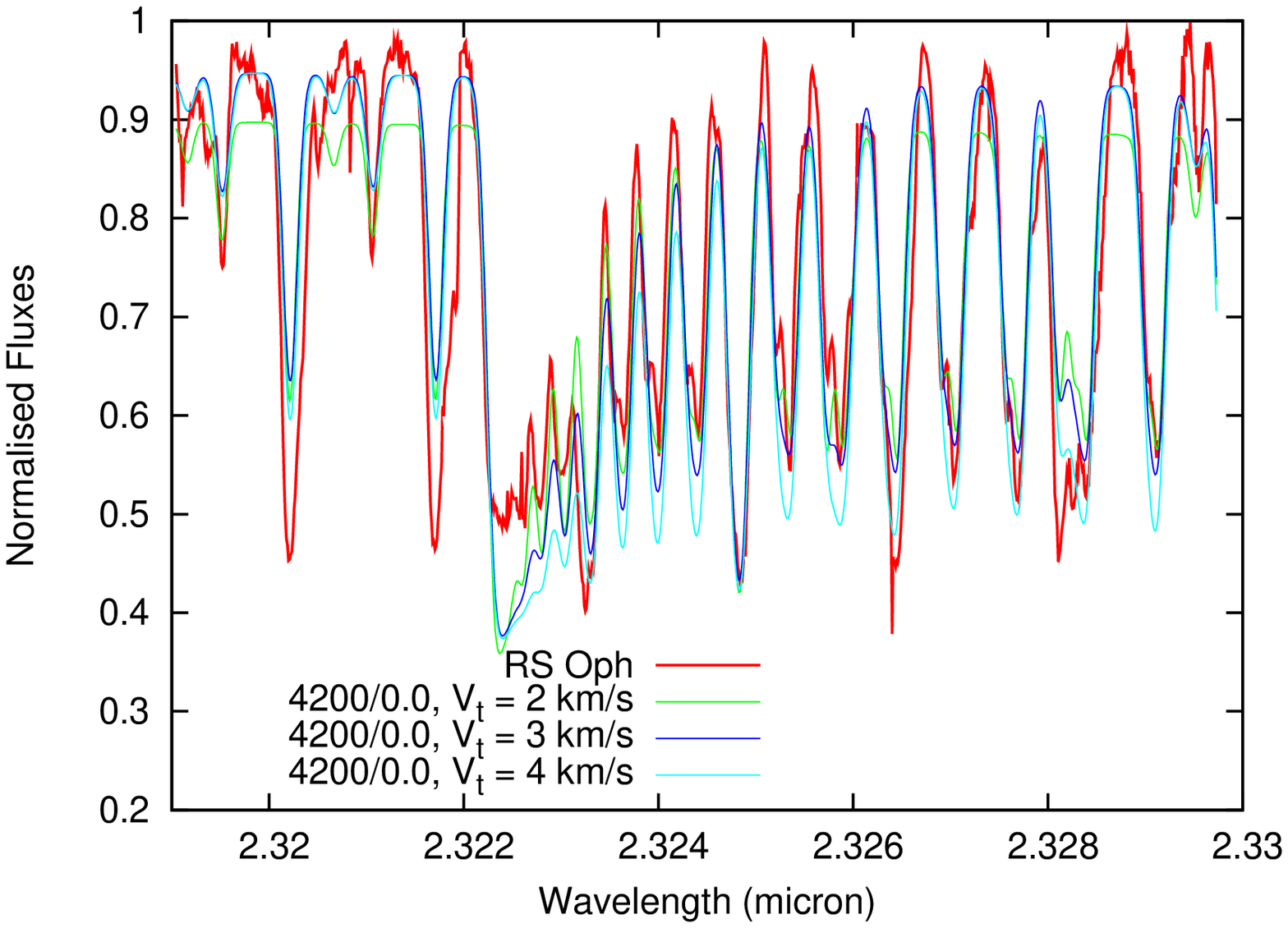}
\caption{Left: dependence of $S$ on the adopted \Vt. 
Right: fits of spectra computed with different \Vt\ to the observed ``B''
spectrum of \RS. Model atmosphere has $T_{\rm eff}=4200~\mbox{K}, \log{g}=0.0$
\citep{rushton09} 
\label{fig:_vt}
}
\end{figure*}

\begin{figure*}
\includegraphics [width=176mm]{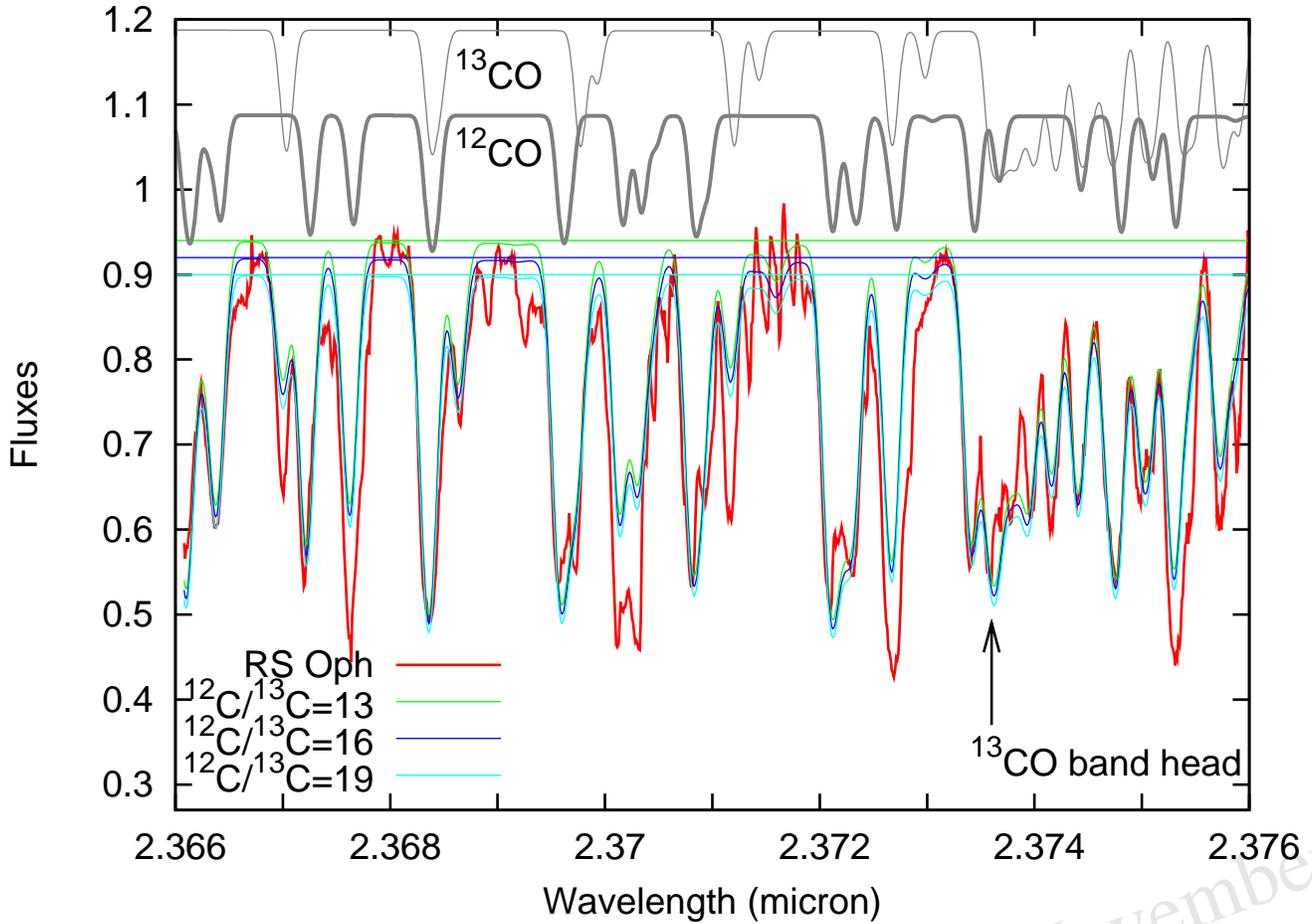}
\caption{Spectral fits computed with different 
\CDC\ to the observed ``R'' spectrum of \RS. 
Pseudontinuum positions at $r_{\nu}$ = 0.90, 0.92, 0.94 are shown; the colour
key is as per the \CDC\ ratio.
Spectra of \CC and \CCO  molecules taken from the Fig.1 are shown on the top.
\label{fig:_cdc}, ones are shifted on y-axies to simplify the plot. 
}
\end{figure*}

\begin{figure}
\includegraphics [width=88mm]{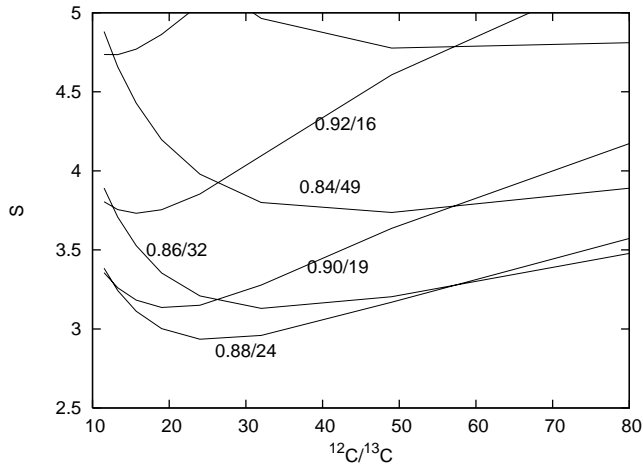}
\caption{
Dependence of $S$ on the 
adopted \CDC. The model atmosphere has $T_{\rm eff} = 4200~\mbox{K}, 
\log{g}=0.0$ \citep{pavlenko08_rsoph}, \Vt~=~3\vunit. Curves on the plot are 
labelled by a pair of numbers {\tt d/xxx}, where {\tt d} and {\tt xxx} 
are the adopted level of continuum and the \CDC\ ratio found at the 
min $S$.
\label{fig:_cdcb} 
}
\end{figure}

To determine the \CDC\ ratio, we fitted computed spectra to the 
observed ``R'' spectrum. We note that: 

\vspace{-2mm}

\begin{enumerate}

\item in the computed spectrum we see flux maxima which can be treated 
as continuum. At these wavelengths the contribution of absorption in 
the line wings is negligible. 

\item the ``R'' spectrum is affected by noise, as already noted. We 
attribute these features to noise because their widths do not match 
the spectral resolution; 

\item to study the effect the continuum level has on our results we 
varied the level of continuum in the observed spectrum; for each 
individual level we found values of \CDC. 

\end{enumerate}

The continuum level in the observed spectrum was adopted at the following 
points: $d = r_{\nu}/r_{max} = 0.88, 0.90, 0.92, 0.94, 0.96, 0.98, 1.00$, 
the value $d= 1.00$ corresponding to the level of the formal maximum flux 
$F_{max}$ in the observed spectrum, here $r_{\nu} = F_{\nu}/F_c$ is the
residual flux. Adopted levels of continuum $F_c$ in the observed spectrum are
clearly seen in Fig.~\ref{fig:_cdc}.

In the Fig.~\ref{fig:_cdc} we show the dependence of the  
synthetic spectra on  the \CDC\ ratio for \Vt=3\vunit, while in
Fig.~\ref{fig:_cdcb} we show  the dependence of the best-fit  parameter $S =
\sum(1-r_{\nu}^{\rm synt}/r_{\nu}^{\rm obs})^2$ on   the \CDC\ ratio for
different levels of continuum $d$  in the observed spectrum. 

Using our fitting procedure, we found a clear dependence of the  minimum $S$
value and of the \CDC\ ratio on the level of pseudo-continuum.  The formal
best-fit (\CDC$=24\pm 3$) is obtained  with $d = 0.88$. However, visual
comparison with the observed  spectrum shows that the level of the theoretical
pseudo-continuum is  too low in this case.  As we see from
Fig.~\ref{fig:_cdc}, the best-fits with \CDC$ = 13 - 19$ well reproduce
the profiles of the CO lines, dependent on the chosen pseudo-continuum and
noting that the values of $S$ are high due to the presence of unidentified
spectral features.
Most likely, these features form beyond the atmosphere of the RG
(see below). Therefore, using goodness of fit as judged ``by eye'' 
to complement
our minimization procedure, we choose the value $\mbox{\CDC} = 16 \pm 3$
from the  best-fits of synthetic spectra and pseudo-continuum, computed for our 
assumed \Tef, $\log{g}$ and abundances to the observed ``R'' spectrum.

Interpretation of the ``R'' spectrum is not as 
straight-forward as that of the ``B'' spectrum, in that the noise 
level increases to longer wavelengths. Indeed the noise level in the 
``R'' spectral region is generally higher than that in the ``B'', likely 
due to a decrease in atmospheric transmission longward of about 2.4\mic. 
Also, absorption features are seen in the ``R'' spectrum which 
cannot be identified with CO; most likely, these are a consequence of the 
data reduction procedure, where time-variable telluric features are not 
adequately removed from the spectra.  Indeed comparison of the computed
and ``R'' spectra reveals the presence of some unidentified features in the
observed spectrum, although some of these may be artifacts produced by the
removal of the telluric spectrum. This is not true for the ``B'' spectrum.
The \CCO\ lines, which are our main 
interest, are clearly seen in the ``R'' spectral region; none of the 
unidentified features in the ``R'' spectrum region are identified with 
\CC, \CCO\ lines and so their presence does not affect our conclusions. 

\section{DISCUSSION}
\label{Discuss}

We have determined the \CDC\ isotopic ratio in the atmosphere of the \RS\ 
secondary for orbital phase 0.78 -- when approximately 88\% of the visible 
RG surface was irradiated by ultraviolet radiation emitted by the WD -- to 
be $16 \pm 3$. This is consistent with that expected  after the first 
dredge-up \citep{charbonnel94}, and suggests that the RG in \RS\ has 
undergone first dredge-up  and has evolved beyond this evolutionary stage. 

In spectra shown by \cite{rushton09}, numerous emission lines are 
visible in the spectral regions around $1.0-1.8$\mic. These form in 
material ejected in the 2006 eruption, or in the shocked or ionized RG 
wind, in each case well away from the photosphere of the secondary. Moreover, 
in optical spectra we see pronounced effects of irradiation. 

By the time of our observations in 2008, \RS\ had essentially returned to 
quiescence, and we do not expect irradiation to have a strong impact on 
the spectrum in the 2\mic\ region. Furthermore, the molecular bands we 
observe in the IR are formed deep in the atmosphere of the RG. However, as 
evidenced by the strong emission line spectrum \citep{rushton09}, a hot 
circumstellar envelope exists and may affect 
the spectral details, even in the IR. As we see in Fig.~\ref{fig:_vt}, 
the fits to the saturated \CC\ $\upsilon=3\rightarrow1$ band at 
$\lambda$~2.32\mic\ are imperfect, in that the observed molecular lines 
are shallower than the theoretical bands, which are more saturated. This 
poorer fit could, in part, be due to the uncertainty in the continuum 
level in the ``R'' spectrum (Fig. \ref{fig:_cdc}). On the other hand, 
weakening of the CO can be interpreted as a consequence of higher 
temperatures in the regions where it is formed, so that the structure of 
the outer layers of the atmosphere of the RG may differ from the classical 
model atmosphere assumed here.

We note that \cite{rushton09} report weaker CO first overtone 
absorption in their 2007 observations than in their 2006 and 2008 observations.
Similar changes were observed following the 1985 outburst by \cite{scott94},
who attributed this behaviour to contamination of the RG by the 1985 ejecta.
\citeauthor{rushton09} conclude that the RG is intrinscially variable.
In addition we may expect strong effects due to inhomogeneities, a stellar wind,
and NLTE, but a consideration of these factors is beyond the scope of this
paper. These phenomena affect the spectra of single canonical giants like
Arcturus (see discussion in \citealt{tsuji09}).

With the mass function from \cite{fekel00}, inclination from
\cite{fekel00} and \cite{ribeiro09}, we determine a mass 
$\sim2.25$\Msun\ for the
RG if the WD is close to the Chandrasekhar limit. While the \CDC\ ratio after
first dredge up is insensitive to the mass of the RG progenitor (and of course
the situation in \RS\ is complicated by its binarity and RN eruptions), the
\CDC\ ratio we find is consistent with this value because, on the basis of the
\CDC\ ratio, we can probably exclude a progenitor mass for the RG below
$\sim1$\Msun\ \citep[cf. Fig.~2 of][]{charbonnel94}. We can also of 
course conclude that the WD progenitor had mass $>2.25$\Msun\ so that some
0.8\Msun\ of material has been deposited in the environment of RS~Oph in
addition to any material contributed by the RG wind.

\section{CONCLUSIONS}
\label{Conc}

We have carried out an analysis of a high resolution spectrum of the RG 
component of the RN \RS. We find \CDC$=16 \pm 3$, which sheds new light on 
the history and evolution of the \RS\ system. The secondary star in \RS\ 
is a RG with a deficiency of carbon and an overabundance of nitrogen, as 
predicted by theories of the first approach to the RGB after first dredge 
up. The \CDC\ ratio we have found is consistent with the typical values 
\citep{smiljanic09} observed in the atmospheres of RG stars in open 
clusters after first dredge up, and further suggests that the RG in \RS\ 
has undergone some further mixing, which would have depressed the \CDC\ 
ratio still further.

Furthermore, in view of the strong variability of the CO bands following 
the 1985 \citep{harrison93,scott94} and 2006 \citep{rushton09} 
eruptions, it is important to monitor the IR spectrum  of the RG in \RS\ 
to understand the reason for the changes observed.  

We will provide a detailed analysis of abundances in the RG in \RS\ in
forthcoming papers.

\section{Acknowledgments}

This manuscript is based on observations obtained at the Gemini Observatory,
which is operated by the Association of Universities for Research in Astronomy,
Inc., under a cooperative agreement with the NSF on behalf of the Gemini
partnership: the National Science Foundation (United States), the Science and
Technology Facilities Council (United Kingdom), the National Research Council
(Canada), CONICYT (Chile), the Australian Research Council (Australia),
Minist\'{e}rio da Ci\'{e}ncia e Tecnologia (Brazil) and Ministerio de Ciencia,
Tecnolog\'{i}a e Innovaci\'{o}n Productiva (Argentina).

This work was supported by an International Joint Project Grant from the UK
Royal Society. YP's and BK's studies are partially supported by a program
Cosmomicrophysics of NASU Ukraine. This research has made use of the SIMBAD
database, operated at CDS, Strasbourg, France.
We thank an anonymous Referee for helpful comments on an earlier version
of this paper.



\end{document}